\documentclass[conference]{IEEEtran}
\IEEEoverridecommandlockouts
\usepackage{cite}
\usepackage{amsmath}
\usepackage[qm]{qcircuit}
\usepackage{braket}
\usepackage{amsmath,amssymb,amsfonts,bm}
\usepackage{algorithmic}
\usepackage{graphicx}
\usepackage{textcomp}
\usepackage{xcolor}
\usepackage{comment}
\usepackage[bookmarks = true, pdfpagemode = None, pdfstartview = FitH, colorlinks = true, urlcolor = blue]{hyperref}

\newcommand\Erase{\bgroup\markoverwith{\textcolor{red}{\rule[.5ex]{2pt}{0.4pt}}}\ULon}
\newcommand\Eraseadd{\bgroup\markoverwith{\textcolor{red}{\rule[.5ex]{2pt}{0.4pt}}}\ULon}
\newcommand{\CenterRow}[2]{
  \dimen0=\ht\strutbox%
  \advance\dimen0\dp\strutbox%
  \multiply\dimen0 by#1%
  \divide\dimen0 by2%
  \advance\dimen0 by-.5\normalbaselineskip%
  \raisebox{-\dimen0}[0pt][0pt]{#2}}

\usepackage{tikz}
\newcommand\copyrighttext{
  \footnotesize \textcopyright 2024 IEEE.  Personal use of this material is permitted.  Permission from IEEE must be obtained for all other uses, in any current or future media, including reprinting/republishing this material for advertising or promotional purposes, creating new collective works, for resale or redistribution to servers or lists, or reuse of any copyrighted component of this work in other works.}
\newcommand\copyrightnotice{
\begin{tikzpicture}[remember picture,overlay]
\node[anchor=south,yshift=10pt] at (current page.south) {\fbox{\parbox{\dimexpr\textwidth-\fboxsep-\fboxrule\relax}{\copyrighttext}}};
\end{tikzpicture}%
}

\def\BibTeX{{\rm B\kern-.05em{\sc i\kern-.025em b}\kern-.08em
    T\kern-.1667em\lower.7ex\hbox{E}\kern-.125emX}}

\begin{document}

\copyrightnotice
\pagestyle{plain}
\title{Electric Power Demand Portfolio Optimization \\by Fermionic QAOA \\with Self-Consistent Local Field Modulation
}

\author{\IEEEauthorblockN{Takuya Yoshioka}
\IEEEauthorblockA{\textit{TIS Inc.} \\
2-2-1 Toyosu, Koto-ku, Tokyo 135-0061, Japan\\
email: yoshioka.takuya@tis.co.jp}
\and
\IEEEauthorblockN{Keita Sasada}
\IEEEauthorblockA{\textit{TIS Inc.} \\
2-2-1 Toyosu, Koto-ku, Tokyo 135-0061, Japan\\
}
\and
\IEEEauthorblockN{Yuichiro Nakano}
\IEEEauthorblockA{\textit{Osaka University} \\
1-2 Machikaneyama, Toyonaka, Osaka 560-0043, Japan\\
}
\and
\IEEEauthorblockN{Keisuke Fujii}
\IEEEauthorblockA{\textit{Osaka University} \\
1-2 Machikaneyama, Toyonaka, Osaka 560-0043, Japan\\
}  
\IEEEauthorblockA{\textit{RIKEN} \\
2-1 Hirosawa, Wako, Saitama 351-0198, Japan\\
}
}

\maketitle

\begin{abstract}

Quantum Approximation Optimization Algorithms (QAOA) have been actively developed, among which Fermionic QAOA (FQAOA) has been successfully applied to financial portfolio optimization problems.
  We improve FQAOA and apply it to the optimization of electricity demand portfolios aiming to procure a target amount of electricity with minimum risk.
  Our new algorithm, FQAOA-SCLFM, allows approximate integration of constraints on the target amount of power by utilizing self-consistent local field modulation (SCLFM) in a driver Hamiltonian.
  We demonstrate that this approach performs better than the currently widely used $XY$-QAOA and the previous FQAOA in all instances subjected to this study.

\end{abstract}


\section{Introduction}\label{sec:intro}


Energy resource aggregation is currently attracting attention \cite{Lu, Albadi}.
This involves the procurement and sale of electricity through demand response (DR),
which changes demand patterns by bundling and controlling the energy resources owned by consumers \cite{Lu, Albadi, Palensky, Siano, Nosratabadi}. 
In this DR, an aggregator requests multiple consumers
who have contracted in advance to save electricity and purchases the demand suppression (negawatt) from DR participants,
which is called ``negawatt trading'' \cite{Okawa, Tushar, Tsurumi1, Tsurumi2}.

Since the electricity usage of DR participants fluctuates depending on times of days,
the amount of negawatt obtained by aggregators also fluctuates on such time periods.
Therefore, if the variance in total negawatt can be minimized by combining the participants subject to DR requests,
the aggregator may be able to procure the desired amount of negawatt more accurately \cite{Lu, Tsurumi1, Tsurumi2, Faia}.
However, this portfolio optimization for DR requests involves solving an integer programming problem,
which is expected to become more difficult,
especially as more renewable energy options become available and
as more consumers participate in the DR market \cite{Tsurumi2, Faia}.

Quantum computation is expected to have the potential to efficiently solve combinatorial optimization problems in such power systems \cite{Ajagekar, Koretsky, Nikmehr}.
In particular, the quantum approximation optimization algorithm (QAOA) \cite{Farhi2}
and algorithms derived from it are still being actively studied \cite{Hadfield1, Bravyi, Blekos}.
Among them, we have developed a fermionic QAOA (FQAOA) for efficiently solving constrained optimization problems in shallow circuits \cite{Yoshioka1, Yoshioka2}.


In this study, we propose a new algorithm FQAOA with self-consistent local field modulation (FQAOA-SCLFM)
using a Hartree-Fock (HF) driver Hamiltonian in the original FQAOA framework \cite{Yoshioka1}.
The algorithm is applied to the electric power demand portfolio optimization problems
to incorporate the balance condition between the desired procured power and total negawatt into the HF driver Hamiltonian.
Simulation results show that our FQAOA-SCLFM can procure negawatt more stably than the conventional $XY$-QAOA \cite{Hadfield1, Wang, Niroula} and our previous FQAOA \cite{Yoshioka1, Yoshioka2}.
In addition, a quantitative comparison of expected cost showed that FQAOA-SCLFM outperforms the previous algorithms in all DR time periods.

This paper is organized as follows.
In section \ref{sec:port}, we propose a cost function for electric power portfolio optimization.
Section \ref{sec:FQAOA-SCLFM} takes the formulation of newly proposed FQAOA-SCLFM.
Section \ref{sec:res} presents simulation results with noiseless environment by using FQAOA-SCLFM.
Section \ref{sec:sum} provides a summary.

\section{Electric Power Demand Portfolio Optimization}\label{sec:port}

Negawatt trading requires an appropriate combination of DR participants to gather negawatt that are close to the desired procurement power and
to ensure that the variance in the total negawatt is as small as possible.
In this section, we first describe the formulation of the electricity demand portfolio optimization according to the Ref. \cite{Tsurumi1, Tsurumi2},
and then explain our newly defined cost function with low computational cost.

In this electric power demand portfolio optimization problem,
the portfolio can be written as a binary bit string ${\bm x}=\{0,1\}^L$,
where $x_l = 1$ $(0)$ to make (or not make) a DR request to participant $l$ in all participants $L$.
The following variance $\sigma_t^2$ at time $t$ is minimized under the supply-demand balance condition:
\begin{align}
  \sigma_t^2 =& E\left[\left(\sum_{l=1}^L p_{t,l}x_l-\sum_{l=1}^L E[p_{t,l}]x_l\right)^2\right]\nonumber\\
  =&\sum_{l=1}^L\sum_{l'=1}^L\sigma_{t,l,l'}x_lx_{l'},\\
\end{align}
s.t. $\quad\sum_{l=1}^L x_l= M$ with the balance condition,
\begin{equation}
  P_{t,{\rm proc}} \le \sum_{l=1}^L E[p_{t,l}]x_l \leq P_{t, {\rm proc}}+\delta,\label{eq:balance}
\end{equation}
where $P_{t,{\rm proc}}$ is the amount of electricity procurement requested by the aggregator,
$p_{t,l}$ is random variable of negawatt that DR participant $l$ can provide,
and $M$ is the number of DR requests submitted.
The $E[\cdots]$ means taking the expected value for the variables $p_{t,l}$'s in parentheses
 and $\sigma_{t,l,l'}=E\left[p_{t,l}p_{t,l'}\right]-E[p_{t,l}]E[p_{t,l'}]$ is the covariance.

In this study, we define an alternative cost function $E_{t,{\bm x}}$ for $\sigma_{t}^2$ as follows:
\begin{align}
  E_{t,{\bm x}}=& E\left[\left(\sum_{l=1}^L p_{t,l}x_l-P'_{t,{\rm proc}}\right)^2\right]\nonumber\\
  =& \sum_{l=1}^L\sum_{l'=1}^L\sigma_{t,l,l'}x_lx_{l'}
  +\left(\sum_{l=1}^LE[p_{t,l}]x_l - P'_{t, {\rm proc}}\right)^2,\label{eq:Etx}
\end{align}
s.t. $\sum_{l=1}^L x_l = M$,
where the inequality constraint indicated by Eq. (\ref{eq:balance}) is introduced in the form of a penalty function in the second term in Eq. ({\ref{eq:Etx}}) with $P'_{t, {\rm proc}} = P_{t, {\rm proc}}+\delta/2$.
This alternative cost function guarantees efficiency and feasibility within the limitations of current quantum hardware while violating the inequality constraint in Eq. (\ref{eq:balance}).

In actual negawatt trading, it is usually executed in a constant portfolio during the time period denoted by $T$.
Therefore, the objective is to find $x^*$ according to the following equations:
\begin{align}
&{\bm x}^* = \arg\min E_{T, {\bm x}}\\
&{\rm for}\quad E_{T, {\bm x}} = \frac{1}{N_T} \sum_{t\in T} E_{t, {\bm x}}, \quad {\rm s.t.} \quad \sum_{l=1}^Lx_l = M_T,\label{eq:costconst}
\end{align}
where $E_{T, {\bm x}}$ is the average value of the cost function $E_{t, {\bm x}}$ shown in Eq. (\ref{eq:Etx}) within time period $T$, where $N_T$ is the number of reference points.

\section{FQAOA with Self-Consistent Local Field Modulation (FQAOA-SCLFM)} \label{sec:FQAOA-SCLFM}
The FQAOA-SCLFM ansatz at QAOA level $p$ takes the following form as in previous FQAOA \cite{Yoshioka1}:
\begin{equation}
  \ket{\psi_p({\bm \gamma},{\bm \beta})}=
  \left[\prod_{j=1}^{p}\hat{U}_m(\beta_j)\hat{U}_p(\gamma_j) \right]\hat{U}_{\rm init}\ket{\rm vac},\label{eq:Ansqaoa}
\end{equation}
with phase rotation $\hat{U}_p(\gamma) = \exp(-i\gamma\hat{\cal H}_p)$ 
and mixing unitary $\hat{U}_m(\beta) = \exp(-i\beta\hat{\cal H}_d)$ 
, where $\hat{\cal H}_p$ and $\hat{\cal H}_d$ are cost and driver Hamiltonian, respectively.

In this study, we propose a new algorithm FQAOA-SCLFM,
where we propose a new HF driver Hamiltonian $\hat{\cal H}^{\rm HF}_d$
and improve the initial state preparation $\hat{U}_{\rm init}$ and mixing unitary $\hat{U}_m$ according to the guidelines in Ref. \cite{Yoshioka1}.
Note that, in general, the following discussion can be applied to problems with additional soft constraints as in the form of Eq. (\ref{eq:Etx}).

\subsection{Cost Hamiltonian and Constraint Operator}

The above constrained optimization problem Eq. (\ref{eq:costconst}) maps to the following minimum eigenvalue problem:
\begin{equation}    
  \hat{\cal H}_{p, T}\ket{\phi_{\bm x}} = E_{T,{\bm x}}\ket{\phi_{\bm x}},\label{eq:Hpeigen}\\
\end{equation}
with
\begin{equation}    
  \hat{C}\ket{\phi_{\bm x}}=M_T\ket{\phi_{\bm x}},\label{eq:qconst}
\end{equation}
where $\ket{\phi_{\bm x}}$ is computational basis states,
$\hat{\cal H}_p$ and $\hat{C}$ are the cost Hamiltonian and constraint operator, respectively.
In the fermionic form, these operators are written as:
\begin{align}
  \hat{\cal H}_{p,T} 
  =&\sum_{l=1}^L\sum_{l'=1}^L \sigma_{T,l,l'} \hat{n}_l\hat{n}_{l'}
  +\frac{1}{N_T}\sum_{t\in T}\left(\hat{P}_t-P_{t,{\rm proc}}'\right)^2,\label{eq:Hp}\\
  \hat{C}=&\sum_{l=1}^L\hat{n}_l,\label{eq:qconstop}  
\end{align}
where $\sigma_{T,l,l'}=\sum_{t\in T}\sigma_{t, l, l'}/N_T$ and $\hat{P}_t$ is total negawatt operator:
\begin{align}
  \hat{P}_t =& \sum_{l=1}^LE[p_{t,l}]\hat{n}_l.
\end{align}
The number operator $\hat{n}_l$ of fermions is defined by $\hat{n}_l = c^\dagger_lc_l$
using the creation (annihilation) operator $\hat{c}^\dagger_l (c_l)$.
The constraint of Eq. (\ref{eq:qconst}) is therefore replaced by a condition of constant number of fermions.

In this fermionic representation,
the computational basis $\ket{\phi_{\bm x}}$ corresponds to the bit string ${\bm x}\in\{0,1\}^L$ with the following form:
\begin{equation}
  \ket{\phi_{\bm x}}=\prod_{l=1}^L\left(\hat{c}^{\dagger}_l\right)^{x_l}\ket{\rm vac}, \label{eq:base}
\end{equation}
with $\hat{n}_l\ket{\phi_{\bm x}}=x_l\ket{\phi_{\bm x}}$,
where the $\ket{\rm vac}$ is a vacuum satisfying $\hat{c}_l\ket{\rm vac}=0$.
Thus, the constrained combinatorial optimization problem is expressed as a problem to find the ground state of
$\hat{\cal H}_{p, T}$ in a subspace with a constant number of particles.

\subsection{Hartree-Fock Driver Hamiltonian}
Here we construct a new efficient driver Hamiltonian satisfying the conditions I, II, and III in Ref. \cite{Yoshioka1}.
From condition I, we assume that $\hat{\cal H}_d$ is a fermion model satisfying the particle number conservation law.
From condition II,  we design $\hat{\cal H}_d$ such that all sites labeled with $l$ are connected by the hopping terms.
From condition III, we can express the ground state of $\hat{\cal H}_d$ as a single Slater determinant satisfying the constraint.

We examine a specific driver Hamiltonian that is adapted to the current problem.
The second term in the cost Hamiltonian of the Eq. (\ref{eq:Hp}) plays an important role like soft constraint between the desired procurement and the total negawatt.
Therefore, we take the following driver Hamiltonian with the same second term in Eq. (\ref{eq:Hp}) as:
\begin{align}
  \hat{\cal H}_{d,T} 
  =&-t_{\rm hop}\sum_{l=1}^L(\hat{c}^\dagger_{l+1}\hat{c}_l + {\rm h.c.})+\frac{1}{N_T}\sum_{t \in T}\left(\hat{P}_t-P'_{t,{\rm proc}}\right)^2,\label{eq:Hd}  
\end{align}
However, because the second term contains a second-order term of the fermion number operator,
its ground state cannot be expressed by a single Slater determinant and does not satisfy condition III.

To overcome this problem, we define the following new HF driver Hamiltonian as:
\begin{align}
  \hat{\cal H}^{\rm HF}_{d,T} 
  =&-t_{\rm hop}\sum_{l=1}^L(\hat{c}^\dagger_{l+1}\hat{c}_l + {\rm h.c.})\nonumber\\
  &+\frac{2}{N_T}\sum_{t\in T}\left(P_{t, {\rm tot}}^{\rm HF}-P'_{t,{\rm proc}}\right)\sum_{l=1}^LE[p_{t,l}]\hat{n}_l\nonumber\\  
  &-\frac{1}{N_T}\sum_{t\in T}\left[(P_{t, {\rm tot}}^{\rm HF})^2-(P'_{t,{\rm proc}})^2\right],\label{eq:HdHF}
\end{align}
with
\begin{equation}
  P_{t,{\rm tot}}^{\rm HF} = \sum_{l=1}^LE[p_{t,l}]\braket{\phi_{\rm HF}|\hat{n}_l|\phi_{\rm HF}}\label{eq:PtHF},
\end{equation}
where $\ket{\phi_{\rm HF}}$ is a ground state of $\hat{\cal H}_{d,T}^{\rm HF}$ in Eq. (\ref{eq:HdHF}) and is determined self-consistently with Eq. (\ref{eq:PtHF}).
The $\hat{\cal H}^{\rm HF}_{d, T}$ is obtained by approximating $(\hat{P}_t-P_{t,{\rm tot}}^{\rm HF})^2 = 0$ for the $\hat{\cal H}_{d, T}$.
The specific iterative process is shown in APPENDIX \ref{sec:iteration}.
We note here that for hopping term in Eq. (\ref{eq:HdHF}) the periodic (anti-periodic) boundary condition is applied for odd (even) number of fermions.

\subsection{Variational Parameter Optimization}
The minimum cost at the QAOA level $p$ is obtained by the following parameter optimization:
\begin{equation}
  E_p(\bm{\gamma}^*,\bm{\beta}^*)=   \min_{\bm{\gamma},\bm{\beta}}
\braket{\psi_p({\bm \gamma},{\bm \beta})|{\cal H}_p|\psi_p({\bm \gamma},{\bm \beta})}.\label{eq:optEp}
\end{equation}
The variational parameters $({\bm \gamma}^*,{\bm \beta}^*)$ are determined by the classical parameter optimization.
This wave function determines the probability $P_{T,{\bm x}}({\bm \gamma}^*,{\bm \beta}^*)$ of observing a bit string $\ket{\phi_{\bm x}}$ as follows:
  \begin{equation}
   P_{T,{\bm x}}({\bm \gamma}^*,{\bm \beta}^*)=\ket{\bra{\phi_{\bm x}}\psi_p({\bm \gamma}^*,{\bm \beta}^*)}|^2.\label{eq:Px} 
  \end{equation}
Using this probability distribution, the expectation values can be evaluated not only in simulation but also in quantum processing unit (QPU) experiments.

\subsection{Implementation on Quantum Circuit}

  In this subsection, we describe how to implement FQAOA-SCLFM ansatz of Eq. (\ref{eq:Ansqaoa})  on quantum circuits.
  The phase rotation unitary $\hat{U}_p$ and initial states preparation unitary
  $\hat{U}_{\rm init}$ can be implemented as in the case of FQAOA \cite{Yoshioka1, Yoshioka2}.
  Therefore, we specifically describe an implementation of mixing unitary
  $\hat{U}_m(\beta) = \exp(-i\beta \hat{H}^{\rm HF}_{d, T})$.

  The unitary $\hat{U}_m(\beta)$ for FQAOA-SCLFM can be written as:
\begin{equation}
  \hat{U}_m(\beta)=\hat{U}_n(\beta)\hat{U}_{\rm BC}(\beta)
  \hat{U}_{\rm odd}(\beta)\hat{U}_{\rm even}(\beta),\label{eq:Trotter}
\end{equation}
with
\begin{align}
  \hat{U}_{\substack{{\rm even}\\{\rm (odd)}}}(\beta)=&\prod_{\substack{l\\{\rm even}\\{(odd)}}}
  \exp\left[i\beta t_{\rm hop}\left(\hat{c}^{\dagger}_l\hat{c}_{l+1}+{\rm h.c.}\right)\right],\label{eq:ophop}\\
  \hat{U}_{\rm BC}(\beta)=&\exp\left[i\beta t_{\rm hop}(-1)^{M-1}\left(\hat{c}^{\dagger}_l\hat{c}_{1}+{\rm h.c.}\right)\right],
  \label{eq:ophopbc}\\
  \hat{U}_I(\beta) =& \prod_{i=1}^L\exp\left(-i\beta I_l\hat{n}_l \right),\label{eq:Un}
\end{align}
where $I_l = (2/N_T)\sum_{t\in T}(P_{t, {\rm tot}}^{\rm HF}-P_{t,{\rm proc}})\sum_{l=1}^LE[p_{t,l}]$
with the convergence value $P_{t, {\rm tot}}^{\rm HF}$ after HF iteration for initial state preparation process (see APPENDIX \ref{sec:iteration}).
The implementation of the unitary operations in Eqs. (\ref{eq:ophop}) and (\ref{eq:ophopbc}) on quantum circuits is given in Refs \cite{Yoshioka1, Yoshioka2}.
The number of gates required for FQAOA ansatz is summarized in TABLE I in Ref. \cite{Yoshioka2}.
The additional unitary $\hat{U}_I$ is implemented by the action of Pauli-$Z$ rotation gates by transformation
$\hat{n}_l\rightarrow (1-\hat{Z}_l)/2$.
As a result, the number of gate operations increased in FQAOA-SCLFM from FQAOA is only $pL$ in Eq. \ref{eq:Ansqaoa}.

\begin{figure}[htb]
  \begin{center}
  \includegraphics[width=8.5cm]{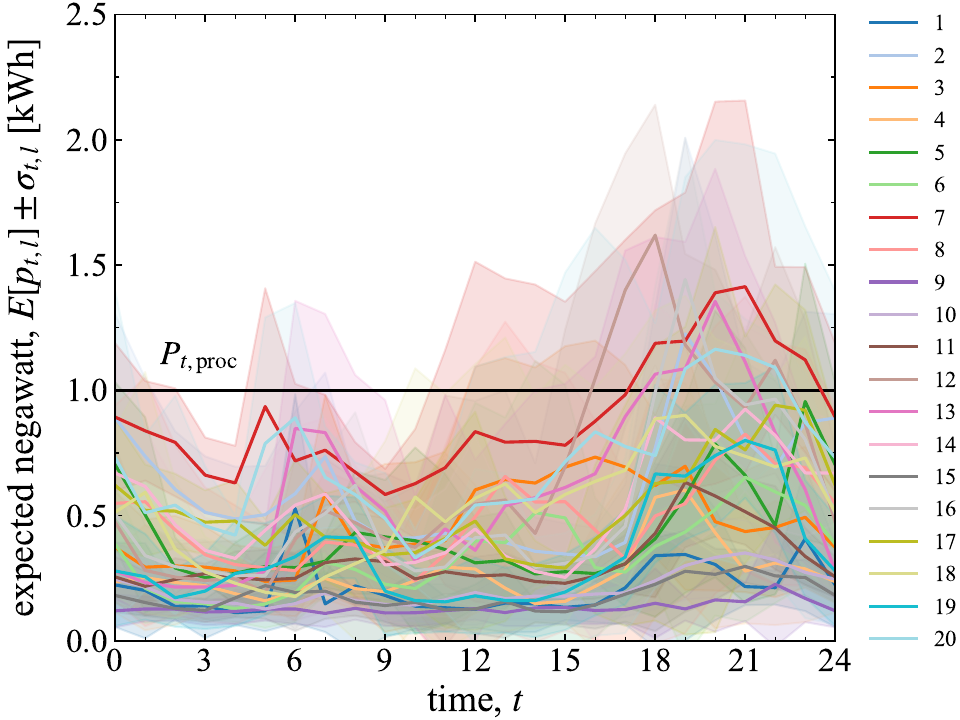} 
  \caption{\label{fig:nW}
    Time series data of expected negawatt $E[p_{t,l}]\pm \sigma_{t,l}$ from DR participants $l=1-20$, where $\sigma_{t,l}$ is standard deviation of
    random variable $p_{t,l}$.
    Here, according to Ref. \cite{Tsurumi1},
    the electricity usage data for each of 20 residences in Ref. \cite{Niigata} is assumed
    to be the amount of negawatt that each participant can supply.
  }
  \end{center}  
\end{figure}
\begin{figure*}[htb]
    \begin{center}
  \includegraphics[width=17cm]{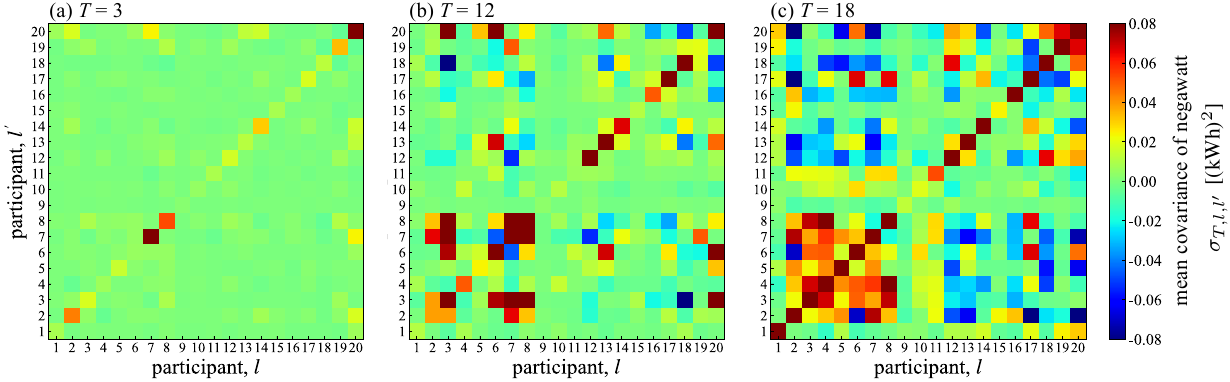} 
  \caption{\label{fig:cov}
    Mean covariance of negawatt $\sigma_{T,l,l'}$ [(kWh)$^2$] in Eq. (\ref{eq:Hp}) between DR participants $l=1-20$ in time periods $T = 3, 12$, and $18$.
    Here we follow the same procedure as in Fig. \ref{fig:nW}. 
  }
    \end{center}
\end{figure*}

\subsection{Computational Details}

In this subsection, we describe the calculation setup.
The range of cost $W_T$ that the $\hat{\cal H}_{p, T}$ can take under the constraint condition
is independent of the type of driver Hamiltonian $\hat{\cal H}_{d, T}$ and initial states,
so we adopt it as the cost scale for performance comparison in time period $T$.
For the model parameter of $t_{\rm hop}$ in Eq. (\ref{eq:HdHF}),
is set so that the cost range of first term in Eq. (\ref{eq:HdHF}) is equal to the first term in Eq. (\ref{eq:Hp}) under the constraints.

Numerical simulations have been performed using the fast quantum simulator qulacs \cite{qulacs}.
The optimized parameters and costs in Eq. (\ref{eq:optEp})
is obtained by the Broyden-Fletcher-Goldfarb-Shanno (BFGS) method.

\section{Results} \label{sec:res}
In this paper, a forecasting model for small-scale electric power demand portfolio optimization is
developed based on historical electricity usage in residences. 
This model is solved with the newly proposed FQAOA-SCLFM.

\subsection{Model Parameters for Electricity Demand Portfolio Optimization}

In this study, we set the time period $T = 3i$ ($0\le i\le 7$) in Eq. (\ref{eq:costconst}), which includes three reference times $t = T, T+1$, $T+2$.
The model parameters $\sigma_{t, l,l'}$ and $E[p_{t,l}]$ at time $t$ in Eqs. (\ref{eq:Etx}) and (\ref{eq:Hp}) are estimated according to Ref. \cite{Tsurumi1} from the database \cite{Niigata}.
Initial test calculations have used $P'_{t,{\rm proc}}=1.5$ kWh in Eqs. (\ref{eq:Hp}) and $M_T=5$ in Eq. (\ref{eq:qconst})
in order to provide a basis for comparing different optimization methods.
Here, the electricity demand of a typical house and a moderate participation rate in DR requests are taken into account.
The model parameters used in this study are shown in Figs. \ref{fig:nW} and \ref{fig:cov}, respectively.

Fig. \ref{fig:nW} shows the expected negawatt $E[p_{t,l}]$ for each DR participant $l$ at time $t$.
It can be seen that at $T = 18$ ($t=$18, 19, and 20) the negawatt with large variance have large time variation.
On the other hand, the variance and time variation of negawatt at midnight $T = 3$ ($t=$3, 4, and 5) are small.

The covariance $\sigma_{T,l,l'}$ among DR participants between $l$ and $l'$ is shown in Fig. \ref{fig:cov}.
Here we show the mean value in time period $T$.
The diagonal component $\sigma_{T, l,l}$ corresponds to the variance of negawatt for participant $l$.
At the midnight (a) $T=3$, the negawatt variance per participant is small.
On the other hand, in time period (c) $T=18$,
the negawatt fluctuation correlations between participants are positively and negatively large.
Thus, we can say that the problem for time period $T=18$ is the difficult to solve.

\begin{figure*}[htb]
    \begin{center}
  \includegraphics[width=17cm]{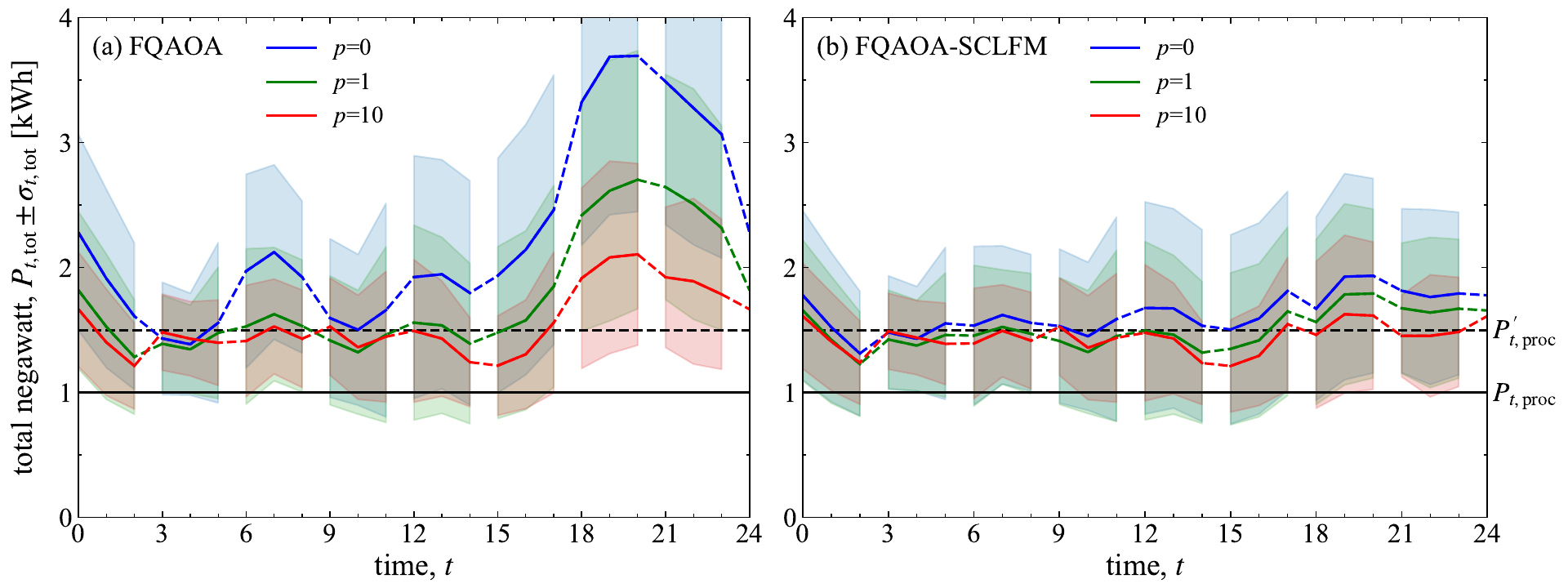} 
  \caption{\label{fig:ts_nW_FQAOA}
    Calculated results of total negawatt $P_{t, {\rm tot}}\pm \sigma_{t, {\rm tot}}$ of Eqs. (\ref{eq:Ptot}) and (\ref{eq:s.d.}) at each time $t$,
    where approximate optimal states $\ket{\psi_p({\bm \gamma}^*, {\bm \beta}^*)}$ are obtained
    from optimization for each time period $T$ with (a) FQAOA and (b) FQAOA-SCLFM at $p=0, 1$, and $10$.
    Procurement of negawatt $P_{t, {\rm proc}}$ and $P_{t, {\rm proc}}'$ are shown by solid and broken lines.
  }
    \end{center}
\end{figure*}

\subsection{results of noiseless simulation}

First, we show the main results of the calculation of the total negawatt expectation and its variance.
Next, we validate the effectiveness of our newly proposed algorithm FQAOA-SCLFM.
For this purpose, we also show the results of calculations with the algorithms $XY$-QAOA \cite{Wang, Niroula} and
the previous FQAOA \cite{Yoshioka1, Yoshioka2} developed in previous studies.
The FQAOA ignores the second term of the driver Hamiltonian of the FQAOA-SCLFM in Eq. (\ref{eq:Hd}).
The $XY$-QAOA replaces the initial state of above FQAOA with a uniform superposition state under the constraint so-called Dicke state \cite{Wang, Aktar, Bartschi}.

\subsubsection{Total Negawatt}

To predict the total negawatt and their variance at time $t$,
we define here the following equations:
\begin{align}
  P_{t, {\rm tot}} =& \braket{\psi_p({\bm \gamma}^*, {\bm \beta}^*)|\hat{P}_t|\psi_p({\bm \gamma}^*, {\bm \beta}^*)}\nonumber\\
  =& \sum_{\bm x}P_{T,{\bm x}}({\bm \gamma}^*, {\bm \beta}^*)\sum_{l=1}^LE[p_{t,l}]x_l, \label{eq:Ptot}
\end{align}  
and
\begin{align}
  \sigma_{t,{\rm tot}}^2
  =&\braket{\psi_p({\bm \gamma}^*, {\bm \beta}^*)|E\left[\left(\sum_{l=1}^Lp_{t,l}\hat{n}_l-P_{t,{\rm tot}}\right)^2\right]|\psi_p({\bm \gamma}^*, {\bm \beta}^*)}\nonumber\\
  =&\sum_{\bm x}P_{T,{\bm x}}({\bm \gamma}^*, {\bm \beta}^*)\sum_{l=1}^L\sum_{l'=1}^LE[p_{t,l}p_{t,l'}]x_lx_{l'}
  -P_{t,{\rm tot}}^2,\label{eq:s.d.}
\end{align}
respectively, where $E[p_{t,l}p_{t,l'}] = \sigma_{t,l,l'}+E[p_{t,l}]E[p_{t,l'}]$.
These incorporate the effects of quantum fluctuations as well as fluctuations in the negawatt prediction.

Fig. \ref{fig:ts_nW_FQAOA} (a) and (b) shows the simulation results of $P_{t, {\rm tot}}\pm \sigma_{t,{\rm tot}}$
of Eqs. (\ref{eq:Ptot}) and (\ref{eq:s.d.}).
By increasing the QAOA level $p$, we can confirm that $P_{t, {\rm tot}}$ asymptotically approaches $P'_{t, {\rm proc}}$ at all time $t$.
However, focusing on the FQAOA results in (a),
the discrepancy between $P_{t,{\rm tot}}$ and $P'_{t,{\rm proc}}$ remains large in the time period $T=18$.
On the other hand, in the case of (b) FQAOA-SCLFM, $P_{t, {\rm tot}}$ is already close to $P'_{t, {\rm proc}}$ at $p=0$, 
This tendency originates from the energy reduction in the HF energy $\braket{\phi_{\rm HF}|\hat{\cal H}_d^{\rm HF}|\phi_{\rm HF}}$,
which is demonstrated in Fig. \ref{fig:histexp} of the APPENDIX \ref{sec:iteration}.
It can also be confirmed that by increasing $p$, the variance decreases significantly, especially in the time periods of $T=18$ and $T=21$.
These results imply that FQAOA-SCLFM outperforms previous FQAOA \cite{Yoshioka1} in all time period $T$.
Finally, we emphasize that as the QAOA level increases to $p=10$,
the total negawatt $P_{t,{\rm tot}}\pm \sigma_{t,{\rm tot}}$ roughly satisfy the balance condition of Eq. (\ref{eq:balance}) and stable procurement of total negawatt is achieved.
\subsubsection{Expected Value of Energy and Probability Distribution}

\begin{table*}[htb]
  \caption{\label{tble:exp}
    Cost expectation $\Delta E_T/W_T$ at $T$ of Eq. (\ref{eq:DE}) calculated by $XY$-QAOA \cite{Wang, Niroula}, FQAOA \cite{Yoshioka1}, and present FQAOA-SCLFM.
  }
  \begin{center}
{\renewcommand\arraystretch{2.0}
  \begin{tabular}{|c c|c|c|c|c|c|c|c|c|}
    \hline
    && \multicolumn{8}{|c|}{$\Delta E_T/W_T$}
    \\ 
    \multicolumn{2}{|c|}{method} &0 &3 &6 &9 &12 &15 & 18& 21\\
    \hline
              & $p=0$   & 0.165 & 0.117 & 0.162 & 0.220 & 0.177 & 0.155 & 0.276 & 0.307 \\
    $XY$-QAOA & $p=1$   & 0.073 & 0.082 & 0.060 & 0.129 & 0.088 & 0.058 & 0.115 & 0.142 \\
              & $p=10$  & 0.033 & 0.037 & 0.024 & 0.053 & 0.033 & 0.021 & 0.052 & 0.056 \\ \hline
              & $p=0$   & 0.168 & 0.123 & 0.162 & 0.222 & 0.177 & 0.156 & 0.275 & 0.305\\
        FQAOA & $p=1$   & 0.066 & 0.082 & 0.056 & 0.123 & 0.076 & 0.051 & 0.096 & 0.121\\
              & $p=10$  & 0.026 & 0.029 & 0.018 & 0.037 & 0.023 & 0.015 & 0.030 & 0.030\\\hline
              & $p=0$   & 0.074 & 0.107 & 0.061 & 0.197 & 0.107 & 0.056 & 0.025 & 0.032 \\  
  FQAOA-SCLFM & $p=1$   & 0.046 & 0.076 & 0.039 & 0.116 & 0.058 & 0.032 & 0.014 & 0.018 \\
              & $p=10$  & 0.018 & 0.027 & 0.017 & 0.036 & 0.019 & 0.012 & 0.007 & 0.007 \\ \hline    
\end{tabular}}
  \end{center}
\end{table*}

Next, to evaluate the performance of our FQAOA-SCLFM quantitatively,
we define the probability distribution of costs $D_T(E)$ and
their expected values $\Delta E_T$ by the following Eqs:
  \begin{equation}
    D_T(E) = \sum_{\bm x}P_{T,{\bm x}}(\bm{\gamma}^*,\bm{\beta}^*)\delta(E-E_{T, {\bm x}}),\label{eq:PE}
  \end{equation}
   and
  \begin{align}
    \Delta E_{T}
    =&\int_{-\infty}^\infty D_T(E)(E-E_{T,{\rm min}})dE\nonumber\\
    =&\sum_{\bm x} P_{T,{\bm x}}(\bm{\gamma}^*,\bm{\beta}^*)
    \left(E_{T,{\bm x}}-E_{T, \rm min}\right),\label{eq:DE}
  \end{align}
  respectively, where $E_{T,{\rm min}}$ is minimum cost at given $T$.
  In illustrating the $D_T(E)$, a finite width integration is performed.

  TABLE \ref{tble:exp} shows the expected cost values simulated by different algorithms with QAOA levels $p=0, 1$, and $10$.
  For any given $T$ and finite $p$,
  FQAOA outperforms $XY$-QAOA, and FQAOA-SCLFM further outperforms FQAOA.
  In contrast to $XY$-QAOA, FQAOAs that incorporating adiabatic time evolution with careful driver Hamiltonian selecton,
  enhancing performance.
The FQAOA-SCLFM gives a more stable initial state due to the HF driver Hamiltonian effectively addressing the penalty in Eq. (\ref{eq:Etx}), as shown in Appendix \ref{sec:iteration}.

\begin{figure}[htb]
    \begin{center}
  \includegraphics[width=8.5cm]{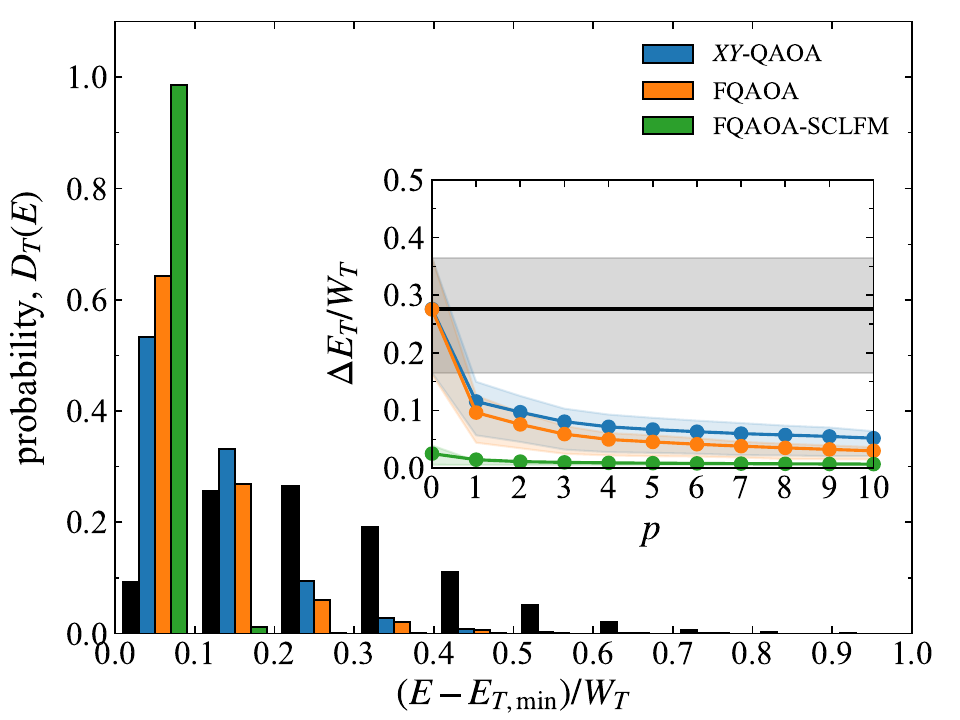} 
  \caption{\label{fig:histsim}
    Probability distribution of cost $D(E)$ at $p=1$ integrated over the $W_T/10$ range of Eq. (\ref{eq:PE}) at $T = 18$.
    Cost expectation values $\Delta E_T/W_T$ of Eq. (\ref{eq:DE}) shown in the inset,
    where we have shaded the region from the first quartile to the third quartile.
    These results are obtained from noiseless simulations using $XY$-QAOA, FQAOA, and FQAOA-SCLFM.
    Black histogram and line in inset indicates the result from random sampling under constraint.}
    \end{center}
\end{figure}

Fig. \ref{fig:histsim} shows the results of the probability distribution for cost obtained by each algorithm at QAOA level $p=1$.
In each case, a peak appears in the lowest energy region.
The peak value is particularly prominent in the FQAOA-SCLFM.
The inset shows the expected cost depending on $p$,
which clearly shows that FQAOA-SCLFM outperforms the other algorithms at all approximation levels $p$.

\section{Summary} \label{sec:sum}

We defined a simple cost function for the electricity demand portfolio optimization problem,
which aims to procure a target amount of power with minimum risk.
We also proposed a new quantum algorithm FQAOA-SCLFM to efficiently solve this problem.
The cost Hamiltonian  consists of a risk term and a penalty term for procuring the target power.
The penalty term is incorporated into the driver Hamiltonian as a self-consistent local field modulation (SCLFM) determined by the Hartree-Fock approximation.
The new algorithm, FQAOA-SCLFM, was shown to outperform conventional $XY$-QAOA and previous FQAOA.
The proposed algorithm is applicable to combinatorial optimization problems of the same form as the present problem in Eq. (\ref{eq:Etx}),
involving both hard and soft constraints.

\appendices
\renewcommand{\theequation}{\thesection.\arabic{equation}}
\thesection
\setcounter{equation}{0}
\section{Hartree-Fock iterations}\label{sec:iteration}

In Hartree-Fock (HF) approximation, Eqs (\ref{eq:HdHF}) and (\ref{eq:PtHF}) are determined self-consistently.
In this study we converge the fermion distribution, which can be written as
\begin{equation}
  n^{\rm HF}_{l,i} = \braket{\phi_{\rm HF}|\hat{n}_l|\phi_{\rm HF}}_i,
\end{equation}
where subscript $i$ is the number of iterations.

\begin{figure}[htb]
    \begin{center}
  \includegraphics[width=8.5cm]{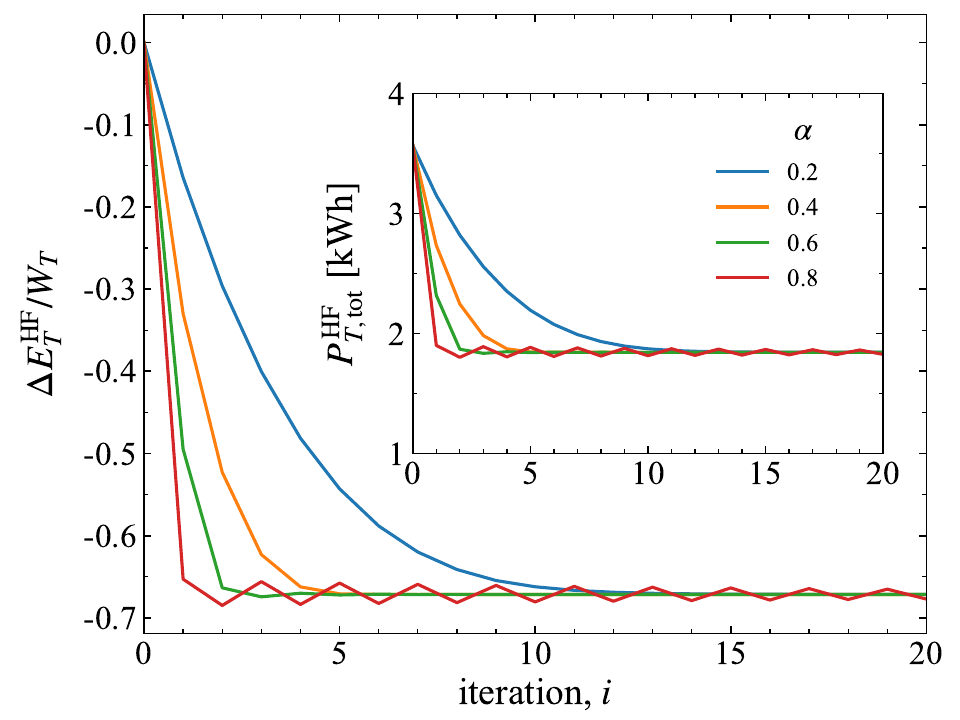} 
  \caption{\label{fig:histexp}
    Decrease of Hartree-Fock energy $E^{\rm HF}_T=\braket{\phi_{\rm HF}|\hat{\cal H}_{d, T}^{\rm HF}|\phi_{\rm HF}}$ using Eq. (\ref{eq:HdHF}) from $i=0$,
    and mean total negawatt
    $P_{T,{\rm tot}}^{\rm HF}=\sum_{t\in T}P_{t, {\rm tot}}^{\rm HF}/N_T$ using Eq. ({\ref{eq:PtHF}}) in time period $T=18$,
    where, $\alpha$ is mixing parameter in Eq. (\ref{eq:mix}).
  }
    \end{center}
\end{figure}

To stabilize the self-consistent iteration,
we introduce mixing parameter $\alpha$ to mix the output fermion distribution $n^{\rm HF (out)}_{l,i+1}$ 
with the old input distribution $n_{l,i}^{\rm HF}$ as:
\begin{align}
  n_{l,0}^{\rm HF} =& 1/M_T, \label{eq:n0}\\
  n_{l,i+1}^{\rm HF} =& (1-\alpha) n_{l,i}^{\rm HF} + \alpha n_{l, i+1}^{\rm HF (out)},\label{eq:mix}
\end{align}
where $n_{l,i+1}^{\rm HF}$ is a new fermion distribution, which be used in the next iteration.
The uniform distribution of $n_{l,0}^{\rm HF}$ in Eq. (\ref{eq:n0}) is that realized in the initial state of $XY$-QAOA and FQAOA.

The HF iteration at $T=18$ for varying the mixing parameter $\alpha$ in equation (\ref{eq:mix}) is shown in Fig. \ref{fig:histexp}.
For $\alpha=0.8$, the energy decrease is fast, however, oscillations appear.
On the other hand, for $\alpha\le 0.6$, we can confirm that the calculation is stabilized.
The inset also shows the expected value of total negawatt during the HF iteration process.

\section*{Acknowledgment}

T.Y. thanks H. Kuramoto for valuable discussions.
K.F. is supported by MEXT Quantum Leap Flagship Program (MEXTQLEAP) Grants No. JPMXS0118067394 and No. JPMXS0120319794. This work is supported by JST COI-NEXT program Grant No. JPMJPF2014.


\end{document}